\documentclass[
 aip,
 jmp,%
 amsmath,amssymb,
 reprint,%
showcase
]{revtex4-1}

\usepackage{graphicx}
\usepackage{dcolumn}
\usepackage{bm}

\newcommand {\pl}[1]{\langle{+,#1}\vert}
\newcommand {\ml}[1]{\langle{-,#1}\vert}

\begin{document}

\preprint{AIP/123-QED}

\title[Manuscript draft]{Trapping photon-dressed Dirac electrons in a quantum
dot studied by coherent two dimensional photon echo spectroscopy }

\author{O. Roslyak, Godfrey Gumbs}
\affiliation{Department of Physics, Hunter College, City University of New York,  \\ 695 Park Avenue, New York, NY, 10065}
\email{avroslyak@gmail.com}
\author{S. Mukamel}
\affiliation{Department of Chemistry, University of California at Irvine, Irvine, CA, 92697}

\date{\today}

\begin{abstract}
We study the localization of  dressed Dirac electrons in a cylindrical
quantum dot (QD) formed on monolayer and bilayer graphene by spatially different
potential profiles. Short lived excitonic states which are too broad to be resolved
in linear spectroscopy are revealed by cross peaks in the photon-echo nonlinear technique.
Signatures of the dynamic gap in the two-dimensional spectra are discussed.
The effect of the Coulomb induced exciton-exciton scattering and
the formation of biexciton molecules are demonstrated.
\end{abstract}

\maketitle

\section{\label{SEC:1} Introduction}

Due to its unique band structure, the charge carriers in graphene are massless Dirac fermions
which can cross high potential barriers with ideal unity transmission coefficient (the Klein paradox)
\cite{castro2009electronic}.
This ensures a very effective escape channel from a trapping potential thus making it hard for conventional Dirac electrons to be localized within graphene based QDs. 
Within a finite spatial region defined by sharp potential profile \cite{xavier2010topological,paananen2011signatures,pal2011electric,park2010tunable,matulis2008quasibound,wunsch2008electron,hewageegana2008electron}.
To overcome this difficulty, and trap the electrons for sufficiently long time,
we propose to dress the electrons with circularly polarized photons, thus providing them with an
effective mass\cite{kibis2010metal,pal2011electric}. 
The localization is demonstrated in a cylindrical quantum dot (QD)
formed in monolayer and bilayer graphene by
antisymmetric potential kink. 
Conventionally the measure of localization are characteristic resonances in the electronic density of states\cite{wunsch2008electron,hewageegana2008electron}.
The dynamical gap is studied semi-classically using Floquet's theorem
\cite{zhou2011optical}.
We present a fully quantum mechanical model, which is  based on dressing electrons in
monolayer and bilayer configurations. 
Our calculations show that the dressing not only opens up
a dynamical gap in the energy dispersion but also renormalizes the Fermi velocity and
inter layer coupling coefficients. 
In the bilayer configuration,
the dressing tunes  the gap. 
That is, it can either close or open the gap, depending on
the polarity of the potential kink and the direction/degree of the polarization.
The resulting confined electronic states should have similarities with the surface states
of topological insulators. 
Their energies are located inside the energy gap and the
wave functions  decay away from the interface of the kink potential. 
These topological states, with the carriers propagating along the potential kink, are expected to be robust
with respect to the effects of disorder \cite{xavier2010topological}.
The fully localized states are mixed with the quasi-bound states above the energy gap
which can effectively carry away the energy.
Conventional linear response spectra (proportional to the density of states)
provides limited information about them due to the large broadening caused by their short lifetime
\cite{hewageegana2008electron}
.
We propose to utilize femtosecond nonlinear spectroscopy in order to study their dynamics.
We shall use a four-wave mixing  technique known as photon-echo\cite{abramavicius2009coherent}.
The mixed signal is heterodyne detected in the direction of
$-\mathbf{k}_1 + \mathbf{k}_2 + \mathbf{k}_3$ as shown in Fig. \ref{FIG:0}.
The photon echo is known to be able to eliminate the inhomogeneous
broadening due to impurities, and allow us to focus on the intrinsic lifetimes of
the electronic states. 
We further discuss signatures of the dynamic gap on the
two-dimensional (2D) spectra.
There is yet another peculiar characteristic of localized Dirac electrons.
As in metals, they are dynamically screened, leading to small
Coulomb interaction between them. 
For small QD this leaves Pauli blocking as the primary source of
the nonlinear signal\cite{wunsch2008electron}. 
This allows us to calculate it as a
sum-over-states (supermolecule) formalism. 
We can further simplify the signal
interpretation by switching to the quasiparticle picture. 
Those are given as
deviation from ideal bosons\cite{abramavicius2009coherent} for which the nonlinear signals vanishes.
We are  able to consider only excited states absorption Liouville pathways
without contribution from the ground state bleaching and excited states emission.
This interference reduction provides relatively simple interpretation of the 2D spectra.
The short lived states can be visualized via the coherences (off-diagonal cross resonances)
with those fully localized. 
We   employ  visible light to map the QD interband
transitions onto the 2D spectra.
Finally we briefly discuss the effect of the Coulomb induced exciton scattering
based on nonlinear exciton equations\cite{kavousanaki2009probing}. 
Possible formation of biexciton molecules is demonstrated.

\par
The outline of this paper is as follows. In Sec.\ \ref{SEC:2}, we present
the model Hamiltonian for graphene irradiated with a circularly polarized electromagnetic field.
Section \ref{SEC:3} is devoted to dressing of electrons in bilayer graphene and a derivation
of the eigenstates.  We deal with the trapping of the dressed electrons within a QD
in Sec. \ref{SEC:4}. Section \ref{SEC:5} presents the absorption and correlation spectrum for dressed electrons in a QD.
We present numerical results in Sec. \ref{SEC:6} and conclude in
Sec. \ref{SEC:6} with a summary of our results.

\section{\label{SEC:2}Dressed electrons in free standing graphene.}

The electronic Hamiltonian of graphene irradiated with an electromagnetic field may be
expressed as\cite{kibis2010metal}:

\begin{gather}
\label{EQ:1_1}
\mathcal{H} = \mathcal{H}_0 + \mathcal{H}_1 + \mathcal{H}_2\\
\label{EQ:1_2}
\mathcal{H}_0 = \hbar \omega_0 a^\dag_0 a_0 + \frac{W_0}{\sqrt{N_0}}\left({\sigma_+ a_0 + \sigma_- a_0^\dag}\right)\\
\label{EQ:1_3}
\mathcal{H}_1 = \hbar v_F \boldsymbol \sigma \cdot \mathbf{k} + \mathcal{I} V(x,y)\\
\label{EQ:1_4}
\mathcal{H}_2 =  \sum \limits_{i =1}^{\infty}
\hbar \omega_i a^\dag_i a_i + \frac{W_i}{\sqrt{N_i}}\left({(\sigma_+ + \sigma_-)(a_i + a^\dag_i)}
\right)  \ .
\end{gather}
Here, $\mathcal{H}_0$ describes the Jaynes-Cummings model\cite{gerry2003introduction} with
$a_0$ being the annihilation operator of a single mode circularly polarized optical field with 
frequency $\omega_0$ and $N_0$ photons in the mode.
Each of them  carries the energy $\hbar \omega_0$.
$\sigma_{\pm} = (\sigma_x \pm i \sigma_y) / 2$ are raising and lowering operators
for $z-$ projection of the electrons pseudo-spin. 
In matrix representation these are Pauli matrices.
$W_0/\sqrt{N_0}$ is the electron-photon coupling, a quantum mechanical analogue of the
classical rotational motion caused by the circularly polarized wave.

\par
$\mathcal{H}_1$ describes conventional Dirac Hamiltonian\cite{castro2009electronic} of graphene with Fermi velocity
$v_F = 10^6 \; m/s$; 
$\mathbf{k}$  is
the wave vector measured from one of the $\mathbf{K}$ points, $V(x,y)$ is an external QD
confining potential; 
$\mathcal{I}$ is the identity matrix.
$\mathcal{H}_2$ describes the rest of the optical modes latter used to probe the dressed states by four-wave mixing process.

\par
The Hamiltonian $\mathcal{H}_0$ may be  diagonalized in a
straightforward way \cite{gerry2003introduction} in the following basis:

\begin{gather}
\label{EQ:1_5}
\vert{\psi_{N_0}}\rangle =
\left({
\begin{array}{c}
\vert{\psi_{+,N_{0}}}\rangle\\
\vert{\psi_{-,N_{0}}}\rangle
\end{array}
}\right) \ , \\
\label{EQ:1_6}
\vert{\psi_{\pm,N_{0}}}\rangle =
\cos \phi \vert{\pm,N_0}\rangle \pm
\sin \phi \vert{\mp,N_0 \pm 1}\rangle\\
\label{EQ:1_7}
\cos \phi = \sqrt{\frac{\Omega_{N_0} + \hbar \omega_0}{2 \Omega_{N_0}}}\\
\notag
\sin \phi = \sqrt{\frac{\Omega_{N_0} - \hbar \omega_0}{2 \Omega_{N_0}}}\\
\label{EQ:1_8}
\Omega_{N_0} = \hbar \omega_0  +  W_0^2 (N_0 +1)/N_0 \ .
\end{gather}
Here, the direct product state $\vert{\pm, N_0 \rangle}$ define the uncoupled electron
with pseudo spin up (+) or down (-) and the optical mode with $N_0$photons.
Eq.~\eqref{EQ:1_6} defines the dressed electron states.
In the basis of Eq.~\eqref{EQ:1_5}, the Jaynes-Cummings Hamiltonian assumes the form

\begin{gather}
\label{EQ:1_9}
\langle{\psi_{N_0}}\vert \mathcal{H}_0 \vert{\psi_{N_0}}\rangle   \\
\notag
=\left({
\begin{array}{cc}
\langle{\psi_{+,N_0}}\vert \mathcal{H}_0 \vert{\psi_{+,N_0}}\rangle
&
\langle{\psi_{+,N_0}}\vert \mathcal{H}_0 \vert{\psi_{-,N_0}}\rangle\\
\langle{\psi_{-,N_0}}\vert \mathcal{H}_0 \vert{\psi_{+,N_0}}\rangle
&
\langle{\psi_{-,N_0}}\vert \mathcal{H}_0 \vert{\psi_{-,N_0}}\rangle
\end{array}
}\right)\\
\notag
=\left({
\begin{array}{cc}
N_0 \hbar \omega_0 + E_g/2 & 0\\
0 & N_0 \hbar \omega_0 - E_g/2
\end{array}
}\right)  \\
\notag
=\mathcal{I} N_0 \hbar \omega_0 + (E_g/2) \sigma_3 \ .
\end{gather}
The first term  is a constant, and may be omitted.

\par
The remaining Hamiltonian matrix elements are calculated in 
Appendix \ref{AP:2}, yielding 

\begin{gather}
\label{EQ:1_10}
\langle{\psi_{N_0}}\vert \mathcal{H}_1 \vert{\psi_{N_0}}\rangle 
= \hbar \tilde{v}_F \boldsymbol \sigma \cdot \mathbf{k} + \mathcal{I} V(x,y)\\
\label{EQ:1_11}
\langle{\psi_{N_0}}\vert \mathcal{H}_2 \vert{\psi_{N_0}}\rangle =  \\
\notag
\sum \limits_{i =1}^{\infty} \mathcal{I}
\hbar \omega_i a^\dag_i a_i + \frac{\tilde{W}_i}{\sqrt{N_i}}
\left({(\sigma_+ + \sigma_-)(a_i + a^\dag_i)}\right) \ ,
\end{gather}
where $\tilde{v}_F \equiv v_F \cos^2 \phi $ is the renormalized Fermi velocity
and $\tilde{W}_i \equiv W_i \cos^2 \phi$ are the renormalized couplings to the probing optical modes.

\par
In  the absence of a potential ($V(x,y) = 0$), the eigenvalues of $\mathcal{H}_0 + \mathcal{H}_1$ are
$\pm \sqrt{(\hbar \tilde{V}_F k)^2 + (E_g/2)^2}$ and the corresponding eigenfunctions are:

\begin{gather}
\label{EQ:1_12}
\Psi_{+} (k) = \texttt{e}^{i \mathbf{k r}}
\left({
\begin{array}{c}
\cos \left({\alpha_k/2}\right)\\
\texttt{e}^{i \beta_k} \sin \left({\alpha_k/2}\right)
\end{array}
}\right)\\
\label{EQ:1_13}
\Psi_{+} (k) = \texttt{e}^{i \mathbf{k r}}
\left({
\begin{array}{c}
\sin \left({\alpha_k/2}\right)\\
-\texttt{e}^{i \beta_k} \cos \left({\alpha_k/2}\right)
\end{array}
}\right)\\
\label{EQ:1_14}
\texttt{Tan} \beta_k = k_y/k_x;\\
\notag
\texttt{Tan} \alpha_k = 2 \hbar \tilde{v}_F k/E_g  \ .
\end{gather}

\section{\label{SEC:3}  Dressed electrons in bi-layer graphene}

\begin{figure}[]
\centering
\includegraphics[width=0.8\columnwidth]{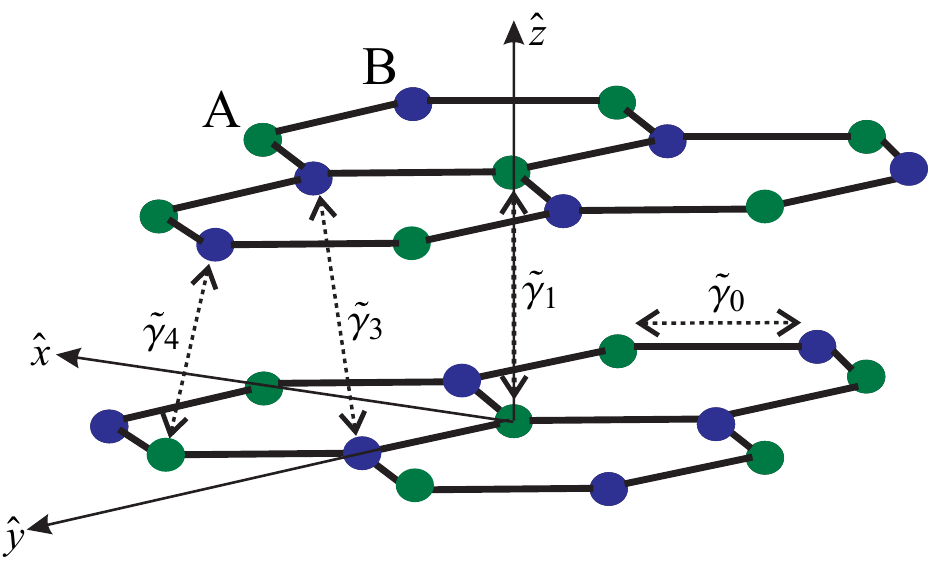}
\caption{\label{FIG:1} Bilayer graphene structure and renormalized coupling coefficients.}
\end{figure}

Starting with Eq. (38) of the review article of Castro Neto, et al. \cite{castro2009electronic}, and applying
the procedure of Appendix \ref{AP:1}, the Hamiltonian which describes the dressing
of the electrons in the bilayer (Bernal stack) can be expressed as

\begin{gather}
\label{EQ:2_1}
H =
\left({
\begin{array}{cc}
H_{11} & H_{12}\\
H_{21} & H_{22}
\end{array}
}\right)
\end{gather}
Here $H_{11} = \mathcal{H} (V_1)$ and $H_{22} = \mathcal{H} (V_2)$
describe the electrons on the first and second graphene layers respectively.
Those layers may experience different potential profiles $(V_{1,2})$ entering Eq.~\eqref{EQ:1_3}.
The interlayer coupling is described by the off-diagonal block matrices as:
\begin{gather}
\label{EQ:2_2}
H_{12} =
\gamma_1 \sigma_{-} + 3 \gamma_3 a (k_x - i k_y) \sigma_{+}  \\
\notag +
\frac{W^3_0}{\sqrt{N_0}} \sigma_{+} a_0
+\sum \limits_{i=1}
\frac{W^3_i}{\sqrt{N_i}} (\sigma_{+} + \sigma_{-}) (a_i + a^\dag_i)\\
\label{EQ:2_3}
H_{21} =
\gamma_1 \sigma_{+} + 3 \gamma_3 a (k_x + i k_y) \sigma_{-}  \\
\notag
+ \frac{W^3_0}{\sqrt{N_0}} \sigma_{-} a^\dag_0
+\sum \limits_{i=1}
\frac{W^3_i}{\sqrt{N_i}} (\sigma_{+} + \sigma_{-} )(a_i + a^\dag_i) \ .
\end{gather}
Here, we have introduced effective electron-photon coupling matrix elements
$W^{j}_{i} /\sqrt{N_i} = - 3 e \gamma_j a \sqrt{2 \pi /\omega_i \Omega \hbar}$,
where $a = 1.42$ \AA~  is the carbon-carbon distance within a layer. Additionally, we have
$\gamma_0 = 2.8 \ , \texttt{eV}$  is the nearest-neighbor
hopping energy within the layer $(A_1 \rightleftharpoons B_1, A_2 \rightleftharpoons B_2 )$.
Fermi velocity can be expressed in terms of the above parameters 
as  $\hbar v _F = 3 \gamma_0 a /2$.
$\gamma_1 = 0.4 \; \texttt{eV}$  is the inter-layer
hopping energy between atoms of type A: $(A_1 \rightleftharpoons A_2)$.
$\gamma_3 = 0.3 \; \texttt{eV}$  is the inter-layer
hopping energy between atoms of type B: $(B_1 \rightleftharpoons B_2)$.
$\gamma_4 = 0.04 \; \texttt{eV}$  is the inter-layer hopping energy 
between atoms of type B and A: $(A_1 \rightleftharpoons B_2, B_1 \rightleftharpoons A_2)$.
Electronic couplings between various atoms in the bilayer graphene are shown in
Fig. \ref{FIG:1}.  The double layer can be regarded as corresponding to
 as $\gamma_1 = \gamma_3 =\gamma_4 =0$.

\par 
In the dressed state basis of Eq.~\eqref{EQ:1_9}, the diagonal 
blocks are given by the results of the previous section as 

\begin{gather}
\label{EQ:2_4}
\langle{\psi_{N_0}}\vert H_{11,22} \vert{\psi_{N_0}}\rangle   \\
\notag
= \hbar \tilde{v}_F \boldsymbol{\sigma} \cdot \mathbf{k} + (E_g/2) \sigma_{3}
+ \mathcal{I} V_{1,2}(x,y) \\
\notag
+ \mathcal{I} N_0 \hbar \omega_0 +
\sum \limits_{i =1} \mathcal{i} \hbar \omega_i a^\dag_i a_i +
\frac{\tilde{W}^0_i}{\sqrt{N_i}} (\sigma_{+} + \sigma_{-}) (a_i + a^\dag_i) \   .
\end{gather}
The off-diagonal blocks may be derived from Eqs.~\eqref{EQ:2_2}, \eqref{EQ:2_3} 
and Appendix \ref{AP:3} to become:
\begin{gather}
\label{EQ:2_5}
\langle{\psi_{N_0}}\vert H_{12} \vert{\psi_{N_0}}\rangle =
\tilde{\gamma}_1 \sigma_{-} + 3 \tilde{\gamma}_3 a (k_x - i k_y) \sigma_{+}  \\
\notag + \tilde{\gamma}_4 \sigma_{3}
+\sum \limits_{i=1}
\frac{\tilde{W}^3_i}{\sqrt{N_i}} (\sigma_{+} + \sigma_{-}) (a_i + a^\dag_i)\\
\label{EQ:2_6}
\langle{\psi_{N_0}}\vert H_{21} \vert{\psi_{N_0}}\rangle =
\tilde{\gamma}_1 \sigma_{+} + 3 \tilde{\gamma}_3 a (k_x + i k_y) \sigma_{-}  \\
\notag
+ \tilde{\gamma}_4 \sigma_{3} +\sum \limits_{i=1}
\frac{\tilde{W}^3_i}{\sqrt{N_i}} (\sigma_{+} + \sigma_{-}) (a_i + a^\dag_i) \ ,
\end{gather}
where the renormalized model  parameters are
$\tilde{\gamma}_{1,2} = \gamma_{1,2} \cos^2 \phi$, $\tilde{\gamma}_4 = (W^3_0 / 2)
\sin 2 \phi$ and
$\tilde{W}^3_i = W^3_i \cos^2 \phi$. For the purpose of further discussion, it is
convenient to localize $H_0 + H_1$ as we did in a preceding section for monolayer
graphene. The corresponding matrix elements are

\begin{widetext}
Single layer:
\begin{gather}
\label{EQ:2_7}
\langle{\psi_{N_0}}\vert
\langle{B_1 A_1}\vert
\mathcal{H}_0 + \mathcal{H}_1
\vert{A_1 B_1}\rangle
\vert{\psi_{N_0}}\rangle \\
\notag
=\left({
\begin{array}{cc}
V_1(x,y) +(E_g/2) &\frac{3}{2} \tilde{\gamma}_0 a (k_x + i k_y) \\
\frac{3}{2} \tilde{\gamma}_0 a (k_x - i k_y) & V_1(x,y) -(E_g/2)
\end{array}
}\right)
\end{gather}
Bilayer:
\begin{gather}
\label{EQ:2_8}
\langle{\psi_{N_0}}\vert
\langle{B_2 A_2 A_1 B_1}\vert
H_0 + H_1
\vert{B_1 A_1 A_2 B_2}\rangle
\vert{\psi_{N_0}}\rangle \\
\notag
=\left({
\begin{array}{cccc}
V_1(x,y) +(E_g/2) & \frac{3}{2} \tilde{\gamma}_0 a (k_x + i k_y) & 
\tilde{\gamma}_4 & 3 \tilde{\gamma}_3 a (k_x - i k_y) \\
\frac{3}{2} \tilde{\gamma}_0 a (k_x - i k_y) & V_1(x,y) -(E_g/2) & 
\tilde{\gamma}_1 & -\tilde{\gamma}_4\\
\tilde{\gamma}_4 & \tilde{\gamma}_1 & V_2(x,y) + (E_g/2) & \frac{3}{2} 
\tilde{\gamma}_0 a (k_x - i k_y) \\
3 \tilde{\gamma}_3 a (k_x + i k_y) & -\tilde{\gamma}_4 & \frac{3}{2} 
\tilde{\gamma}_0 a (k_x + i k_y) & V_2(x,y) -(E_g/2)
\end{array}
}\right) \ .
\end{gather}
\end{widetext}
This implies that the dressing of the Dirac electrons in bilayer gives

\begin{figure*}[]
\centering
\includegraphics[width=\textwidth]{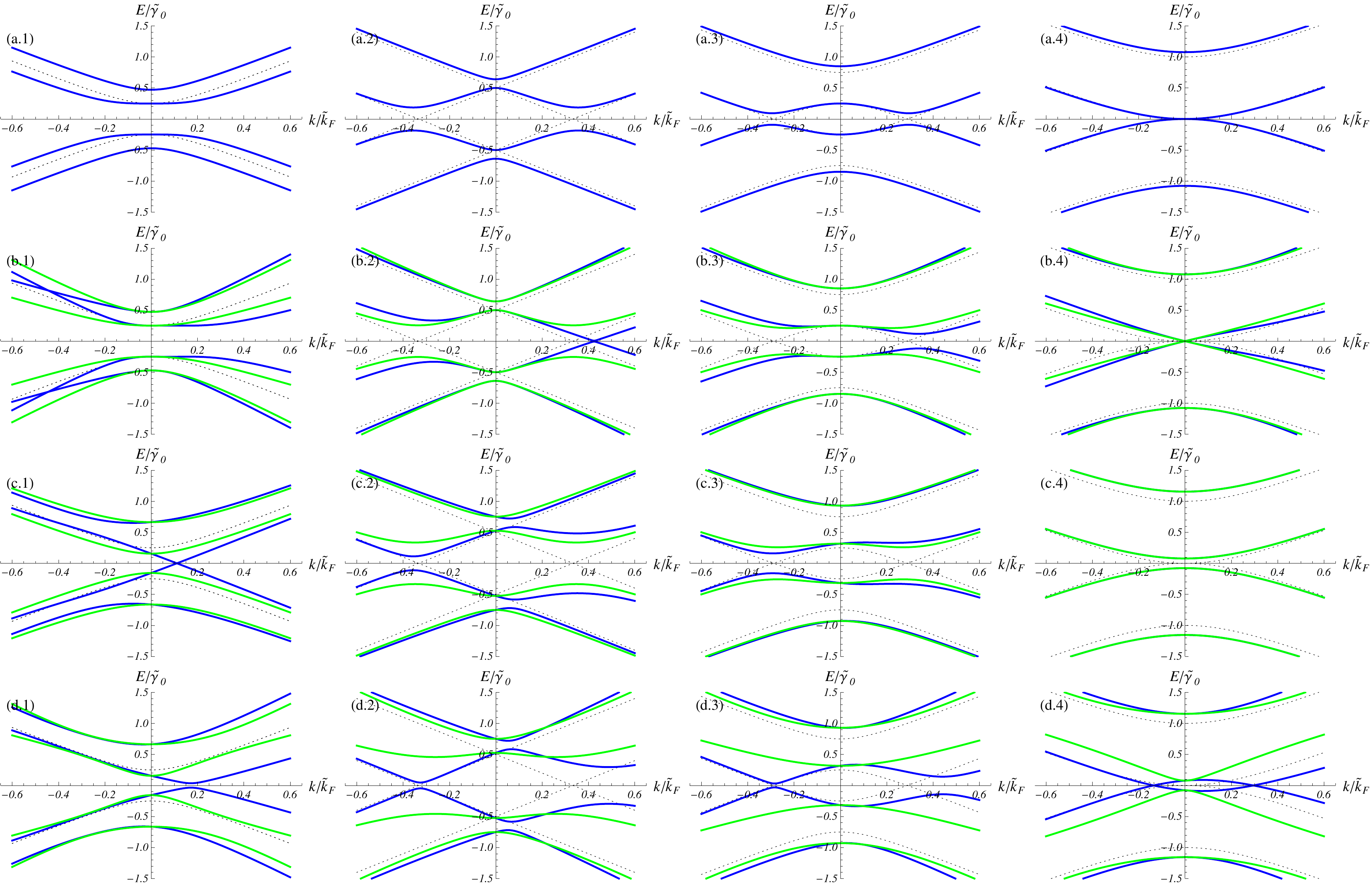}
\caption{\label{FIG:2} Dotted/solid curves represent electron dispersion of
double/bilayer graphene. Panels (a) assume that $\tilde{\gamma}_1/\tilde{\gamma}_0 = 0.4$
and the remaining interlayer coupling parameters are zero.
Panels (b) also introduce the effect of $\tilde{\gamma}_3/\tilde{\gamma}_0 = 0.3$.
Panels (c) demonstrate  the effect of $\tilde{\gamma}_4/\tilde{\gamma}_0 = 0.3$.
Panels (d) illustrate the combined effect of $\tilde{\gamma}_3$ and $\tilde{\gamma}_4$.
In columns (1) we have dressing induced $E_g/\tilde{\gamma}_0 = 1/2$ with no potential
difference between the layers. In columns (2) we potential difference between the layers  $(V_1-V_2)/\tilde{\gamma}_0 = 1/2$ with no electron dressing $E_g/\tilde{\gamma}_0 = 0$.
Columns (3,4) illustrate their combine effect; (3) correspond to
$E_g/\tilde{\gamma}_0 = 1/2$ and (4) has $E_g/\tilde{\gamma}_0 = 1$.
Blue (green) curves show a section of the energy along $k_x (k_y)$ directions.}
\end{figure*}

\begin{itemize}
\item{ renormalized interlayer coupling coefficients, which are denoted by tilde.},

\item{ broken the symmetry between the sub-lattices $(A_1, B_1; A_2,B_2)$ of each 
of the layers. Measure of the broken symmetry is $(E_g/2)$},

\item{ broken symmetry between the sub-lattices $(A_1, B_2; A_2,B_1)$ 
belonging to different layers. A measure of the broken symmetry is $\tilde{\gamma}_4$}.
\end{itemize}
The corresponding eigenvalues for constant potentials ($V_1, V_2$) are shown in
Fig.~\ref{FIG:2} for chosen values of the parameters.
We first focus on the largest interlayer coupling $\tilde{\gamma}_1$ and
neglect the rest of the coupling (Figs. \ref{FIG:2}(a)).
The four bands are given by 

\begin{widetext}

\begin{gather}
\label{EQ:2_9}
\left({2 E + V_2 + V_1}\right)^2  \\
\notag
= E_g^2 + \left({V_1 - V_2}\right)^2 +
9 a^2 k^2 \tilde{\gamma}^2_0 + 2 \tilde{\gamma}_1^2
\pm 2 \sqrt{
\left({V_1 - V_2}\right)^2 \left({E^2_g + 9 a^2 k^2 \tilde{\gamma}^2_0}\right)+
\tilde{\gamma}^2_1 \left({9 a^2 k^2 \tilde{\gamma}_0^2 - 2 E_g \left({V_1 - V_2}\right)}\right)+
\tilde{\gamma}^4_1} \ .
\end{gather}
\end{widetext}
On its own,  $E_g$ opens a gap in the bilayer spectrum similar to the monolayer
(Fig. \ref{FIG:2}(a.1)). The gap may be opened by applying a potential difference
between the layers ($V_1 \neq V_2$ in Fig. \ref{FIG:2}(a.2)).
The combined effect of the potential difference and $E_g>0$ can either widen
$V_2-V_1<0$ or shrink $V_2 - V_1>0$ the gap compared with
the gap induced by the potential difference itself (Fig. \ref{FIG:2}(a.3)).
We observe that when $2 E_g = V_2 - V_1$, the gap closes (Fig. \ref{FIG:2}(a.4)).
Inclusion of the rest of the coupling breaks the symmetry between $k_x$ and
$k_y$, as follows from Fig. \ref{FIG:1}. The analytical form of the energy 
bands, although possible, is too large to be presented here.
The energy bands are shown in Fig. \ref{FIG:2}(b-d).

\section{\label{SEC:4} Dressed electrons confined in a QD }

Let us now turn to the problem of trapping dressed electrons within a QD.
Since the confining potential $V_1(x,y)$ is radial it is convenient to work
with cylindrical coordinates $x = r \cos \theta \; y = r \sin \theta$.
This ammounts to the following substitutions:

\begin{gather}
\label{EQ:2_10}
k_x = - i \partial_x, \; k_y = - i \partial_y\\
\notag
\partial_x \pm i \partial_y = \texttt{e}^{\pm i \theta} 
\left({\partial_r \pm \frac{i}{r} \partial_\theta}\right) \ .
\end{gather}

Thanks to the potential radial symmetry the Hamiltonians
in Eq. \eqref{EQ:2_7} and \eqref{EQ:2_8} commutes with
the angular momentum operator $\hat{L}_z = \mathcal{I} \left({x k_y -y k_x }\right)$.
Here we have neglected symmetry breaking contributions to the Hamiltonian ($\tilde{\gamma}_{3,4} \ll \tilde{\gamma}_{1} $).
 Therefore, we may seek   solution of the Dirac equation in the form: 

\begin{equation}
\label{EQ:2_11}
\vert{\Psi_{m}(\mathbf{r})}\rangle=
\left({
\begin{array}{c}
\psi_{m,B_1} (r) \texttt{e}^{i (m+1/2) \theta}\\
i \psi_{m,A_1} (r) \texttt{e}^{i (m-1/2) \theta}\\
i \psi_{m,A_2} (r) \texttt{e}^{i (m-1/2) \theta}\\
\psi_{m,B_2} (r) \texttt{e}^{i (m+1/2) \theta}
\end{array}}\right)\ .
\end{equation}
Here, the projection of the angular momentum has eigenvalues
 $m=\pm1/2,\pm3/2,\pm5/2,\ldots$. Substituting Eqs. \eqref{EQ:2_11} and \eqref{EQ:2_10} 
 into \eqref{EQ:2_8}, we obtain the following system of ordinary differential equations 
 
\begin{figure}[]
\centering
\includegraphics[width=\columnwidth]{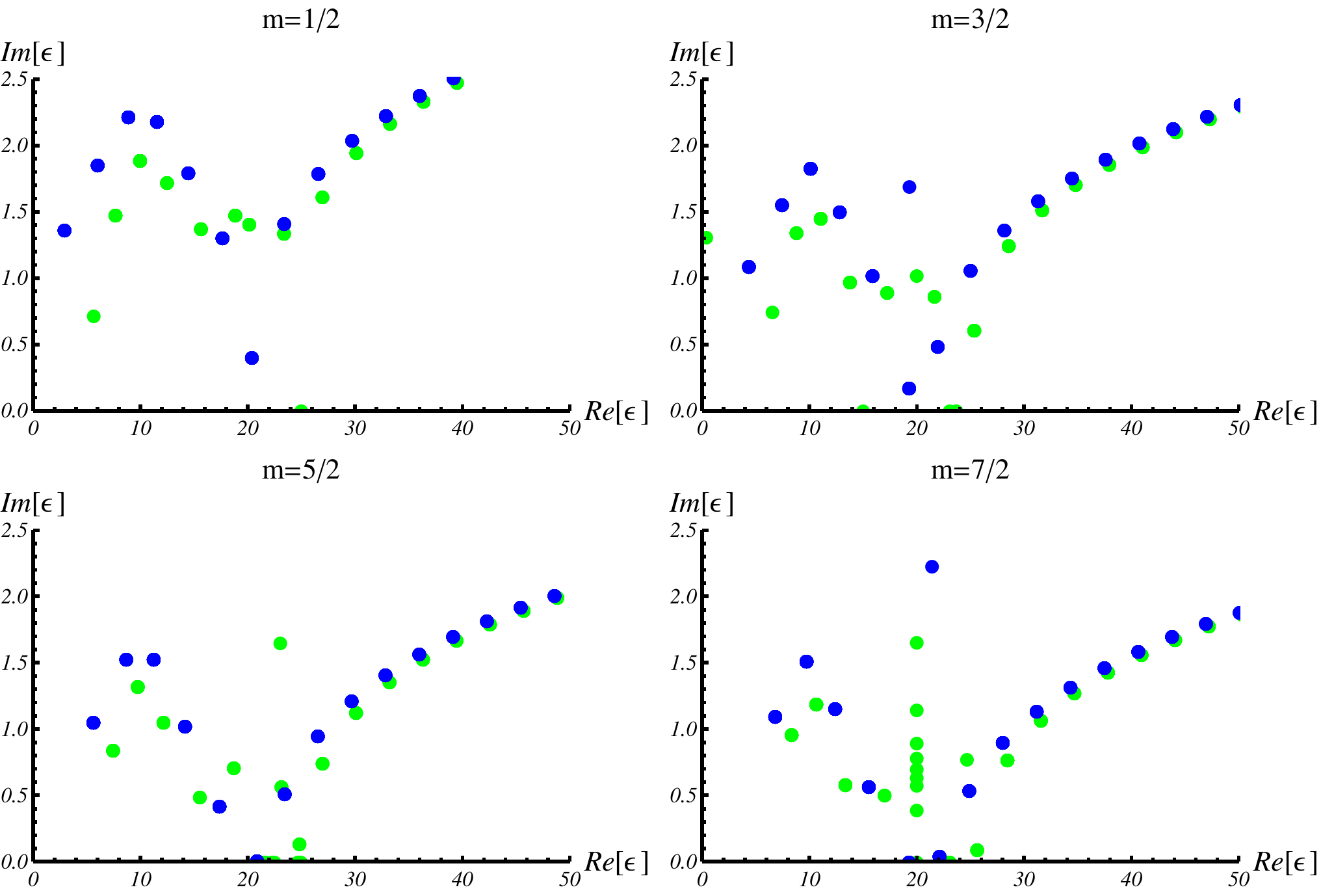}
\caption{\label{FIG:3} Energy levels (Blue dots) of a QD formed in a graphene monolayer.
Green dots correspond to dressed Dirac electrons. The confining potential is
$V_1=20$, and the gap $E_g = 10$ (in units of $2 R/3 \tilde{\gamma}_0 a$).}
\end{figure}

\begin{widetext}

\begin{equation}
\label{EQ:2_12}
\left({
\begin{array}{cccc}
V_1(r) +(E_g/2) & \frac{3}{2} \tilde{\gamma}_0 a (\partial_r - \frac{m-1/2}{r}) & 0 & 0 \\
-\frac{3}{2} \tilde{\gamma}_0 a (\partial_r + \frac{m+1/2}{r}) & V_1(r) -(E_g/2) & \tilde{\gamma}_1 & 0\\
0 & \tilde{\gamma}_1 & V_2(r) + (E_g/2) & -\frac{3}{2} \tilde{\gamma}_0 a (\partial_r + \frac{m+1/2}{r}) \\
0 & 0 & \frac{3}{2} \tilde{\gamma}_0 a (\partial_r - \frac{m-1/2}{r}) & V_2(r) -(E_g/2)
\end{array}
}\right)
\left({
\begin{array}{c}
\psi_{m,B_1}(r)\\
\psi_{m,A_1}(r)\\
\psi_{m,A_2}(r)\\
\psi_{m,B_2}(r)
\end{array}
}\right) = E \left({
\begin{array}{c}
\psi_{m,B_1}(r)\\
\psi_{m,A_1}(r)\\
\psi_{m,A_2}(r)\\
\psi_{m,B_2}(r)
\end{array}
}\right) \   .
\end{equation}
\end{widetext}
First, let us consider the case when there is  no coupling between the graphene layers.
Assuming that  the solution of Eq.  \eqref{EQ:2_12} has the form of $\sqrt{r}\psi_{m}(r)$
in the regions of   constant potential, we obtain 

\begin{gather}
\label{EQ:2_13}
\left({
\begin{array}{c}
\psi_{m,B_1}(r_<)\\
\psi_{m,A_1}(r_<)\\
\end{array}
}\right)= A
\left({
\begin{array}{c}
J_{\vert{m-1/2}\vert}\left({\frac{2 r_< \sqrt{E^2 -(E_g/2)^2}}{3 \tilde{\gamma}_0 a}}\right)\\
J_{\vert{m+1/2}\vert}\left({\frac{2 r_< \sqrt{E^2 -(E_g/2)^2}}{3 \tilde{\gamma}_0 a}}\right)\\
\end{array}
}\right)\\
\label{EQ:2_14}
\left({
\begin{array}{c}
\psi_{m,B_1}(r_>)\\
\psi_{m,A_1}(r_>)\\
\end{array}
}\right)=
B \left({
\begin{array}{c}
H^{(1)}_{\vert{m-1/2}\vert}\left({\frac{2 r_> \sqrt{(E-V_1)^2 -(E_g/2)^2}}{3 \tilde{\gamma}_0 a}}\right)\\
H^{(1)}_{\vert{m+1/2}\vert}\left({\frac{2 r_> \sqrt{(E-V_1)^2 -(E_g/2)^2}}{3 \tilde{\gamma}_0 a}}\right)\\
\end{array}
}\right) \ .
\end{gather}
The Bessel function form of the wave function inside of the QD ($V_1=0, r = r_< \leq R$) 
is dictated by the fact that the wave function must stay finite at $r=0$.
Outside the QD ($V_1>0, r = r_> \geq R$) the wave function must describe the outgoing wave at large distances ($r_> \gg R$).
We, therefore, took it to be the Hankel function of first kind.
At the boundary of the dot the wave function must be continuous.
The energies $E_{m,n}$ of the quasi-stationary states inside of the QD
are obtained by solving the following equation:
\begin{gather}
\label{EQ:2_15}
\frac{H^{(1)}_{\vert{m-1/2}\vert}\left({k_> R}\right)}
{H^{(1)}_{\vert{m+1/2}\vert}\left({k_> R}\right)} 
=\frac{J_{\vert{m-1/2}\vert}\left({k_< R}\right)}
{J_{\vert{m+1/2}\vert}\left({k_< R}\right)} \ ;
\end{gather}
where we have introduced following notation:
\begin{gather*}
k_> = \frac{2 R \sqrt{(E-V_1)^2
-(E_g/2)^2}}{3 \tilde{\gamma}_0 a}\\
k_< = \frac{2 R \sqrt{E^2 -(E_g/2)^2}}{3 
\tilde{\gamma}_0 a}
\end{gather*}
Those are shown in Fig. \ref{FIG:3} for several chosen values of $E_g$.
The long living solutions $\texttt{Im}[E] \approx 0$ can be obtained analytically
by noticing the following identities for the Hankel function in the limit $z \ll 1$,

\begin{gather}
\label{EQ:2_16}
H^{(1)}_n (z) = J_n (z) + i Y_n (z)  \\
\notag
= \frac{1}{\Gamma(n+1)} \left({\frac{z}{2}}\right)^n -
i\frac{\Gamma (n)}{\pi} \left({\frac{2}{z}}\right)^n \ .
\end{gather}
It is clear from the above equation that when $E=V_1 \pm E_g/2$ the left hand side
of Eq. \eqref{EQ:2_15}  vanishes. Therefore the real energies of the QD correspond
to the zeros of the Bessel function with

\begin{equation}
\label{EQ:2_17}
J_{\vert{m-1/2}\vert}\left({\frac{2 R \sqrt{E_{m,n}^2 -(E_g/2)^2}}{3 
\tilde{\gamma}_0 a}}\right) =0 \ .
\end{equation}
The splitting of the central peak in the density of   states (DOS) by the electron 
dressing should be readily accessible in   optical experiments.
This will be the subject of the following section.

\par
Now let us briefly comment on the QD in the bilayer graphene. Since the dressing of the
electrons in the bilayer allows wide manipulation of the gap (See Fig. \ref{FIG:2}(a.3))
one can dynamically form conventional QDs. That is those which support infinitely living
electronic states. There are two possible schematics shown in Fig.
In one case the dressing and the substrate induced gap work concurrently and the whole realization
of the QD is almost identical to the one we had discussed for the single layer.
The only difference is the non-homogeneous potential which forms the dot is the potential between the graphene layers.
Such QD can be readily realized by screening of the substrate potential inside of the QD.
In the other schematic, the combination of the dressing and the substrate potential
closes the gap inside of the QD. The electrons become trapped by the gap outside of the QD.
In both cases the analytical solution of Eq. \eqref{EQ:2_12} may be obtained by following the procedure outlined in Ref.\cite{matulis2008quasibound,xavier2010topological}.

\section{\label{SEC:5}  Absorption and photon-echo spectra of
 dressed Dirac electrons in a single QD }

\begin{figure}[]
\centering
\includegraphics[width=\columnwidth]{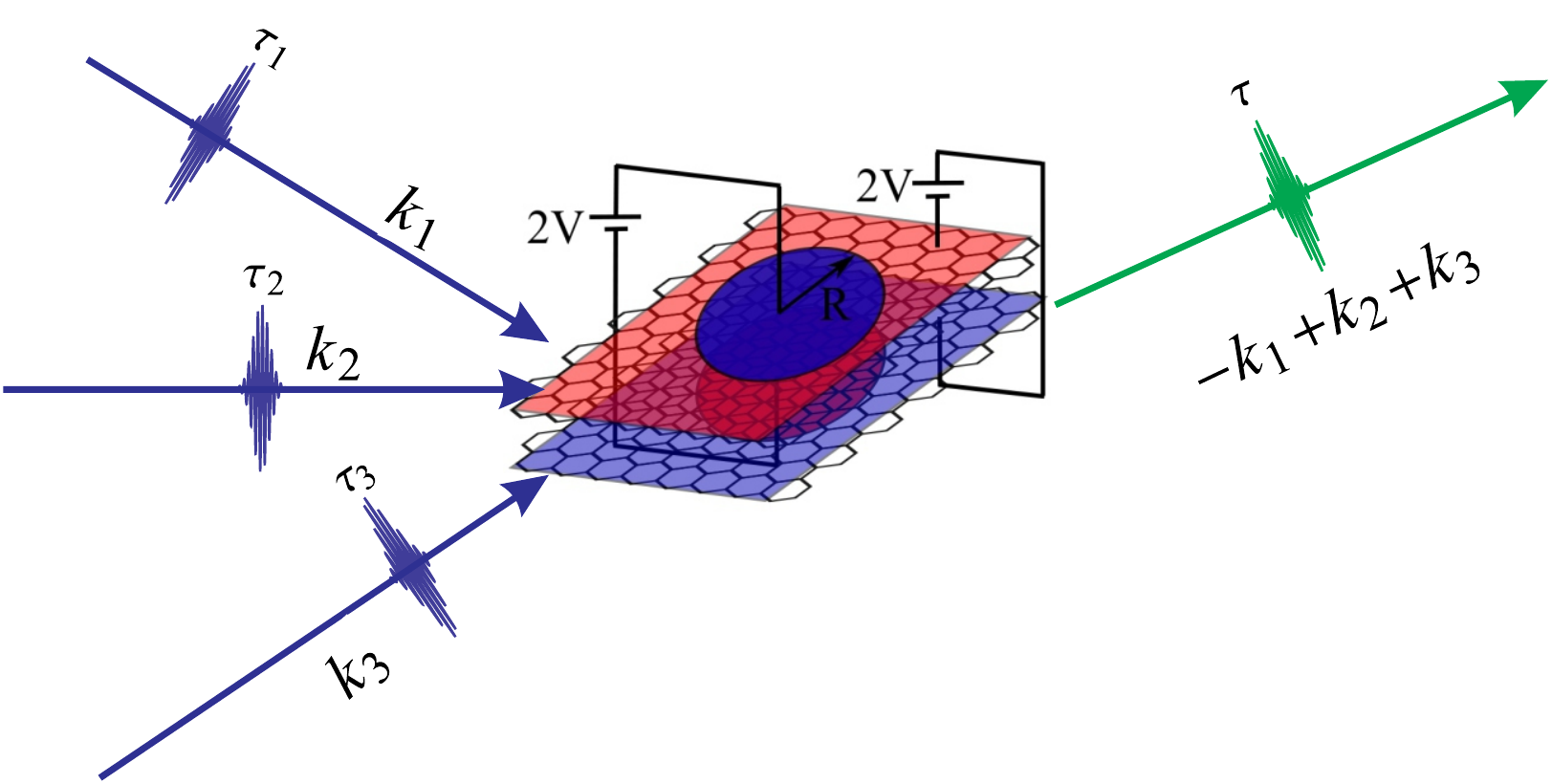}
\caption{\label{FIG:0} Schematic illustration of the photon-echo technique designed
to study the exciton scattering dynamics in graphene based QDs.}
\end{figure}

In this section we investigate several linear and non-linear optical techniques which allow to probe the
details of the electronic structure calculated above.
The linear absorption mostly reveals the long living states ($\texttt{Im}[E] \ll \texttt{Re}[E]$) which
show up a narrow resonance and
directly reflects the structure of the DOS\cite{hewageegana2008electron,wunsch2008electron}.
Nonlinear photon-echo\cite{mukamel1995principles} signal ( $\chi^{(3)}(-\mathbf{k}_1 + \mathbf{k}_2 + \mathbf{k}_3)$) will be designed to reveal the other short living states.
The schematic of the heterodyne detected four wave mixing experiment is shown in
Fig. \ref{FIG:0}
By probing
the coherence between the electronic states in the QD, the technique can reveal the
energy of the short living electronic quasi-bound  states. 
We
shall restrict the discussion to singly excited states, thereby neglecting underlying
many-body effects.  
This allows for a conceptually simple description in terms of the many body eigenstates\cite{abramavicius2009coherent,mukamel2007sum}. 

\par
It is also possible to study several excitonic states of the QD and their dynamics by applying special nonlinear measurements.
We shall also make use of the double quantum coherence\cite{mukamel2007sum} ($\chi^{(3)}(\mathbf{k}_1 + \mathbf{k}_2 - \mathbf{k}_3)$) technique in order to observe many body effects in biexciton manifold of graphene QD.
Conceptually the approach is similar to that described in Ref.\cite{chernyak1998multidimensional,kavousanaki2009probing}. 
However the graphene based QD has large screening of the Coulomb interaction between excitons.
Thus it can be safely omitted in the following discussion.
Recently the collinear version of $\chi^{(3)}$  technique based on phase-cycling gained popularity in QD studies of terahertz regime\cite{kuehn2011strong}.
Its application to graphene will be reported elsewhere.

\par
Let us first define the effective single particle diagonalized Hamiltonian in the QD\cite{mukamel2007sum}:
\begin{gather}
\label{EQ:2_18}
\hat{\mathcal{H}} =
	\sum \limits_{m1,n1} E_{m1} \delta_{m1,n1} \hat{c}^\dag_{m1} \hat{c}_{n1} \\
		\notag
	+\sum \limits_{m2,n2} E_{m2} \delta_{m2,n2} \hat{d}^\dag_{m2} \hat{d}_{n2}
\end{gather}•
whose matrix elements are obtained from Eq.~\eqref{EQ:2_15}.
Each subscript is a composite of two indices describing
angular ($m$) and radial ($n$) quantum numbers $m1(2) =\left\{{m,n}\right\}$.
Here we have partitioned the electronic states into occupied ($n<0$) and unoccupied ($n>0$) in the
ground state obtained by setting up the
chemical potential to $\mu = N_0 \hbar \omega_0$.
The electrons in the unoccupied state can be created by action of $\hat{c}^\dag_{m1} =
\vert{\Psi_{m,n>0}}\rangle \otimes \langle{\Psi_{m,n<0}}\vert$
operator on the ground state.
Its hermitian conjugate $\hat{c}_{m1}$ removes the electron from that state.
Similarly the second term of Eq. ~\eqref{EQ:2_18} describes the creation
$\hat{d}^\dag_{m2} =
\vert{\Psi_{m,n<0}}\rangle \otimes \langle{\Psi_{m,n>0}}\vert$ and annihilation $\hat{d}_{m2}$
of the electrons in the originally occupied states (holes).
The second term in Eq. ~\eqref{EQ:2_18} is just hermitian conjugate of the first.
In the notation above symbol $\otimes$ stands for element-wise multiplication of the vectors.
Since the dressing circularly polarized CW mode has been already incorporated in Eq. ~\eqref{EQ:2_18}
we only have to explicitly treat the interaction
with the narrow pulsed time ordered incoming and detected modes.
This is given by the effective\footnote{The interaction is modified by the dressing, as indicated by tilde.} 
interaction Hamiltonian in rotating wave approximation:

\begin{gather}
\label{EQ:2_19}
\hat{\mathcal{H}}_{\texttt{int}}(t)  
= \sum \limits_{i,m1,m2}
\sum \limits_{j =\pm}
\mathcal{E}^j_i \delta(t-t_i) \mu^j_{m1,m2} \hat{d}_{m2} \hat{c}_{m1} +\texttt{H.c}
\end{gather}•
Here $\mathcal{E}^{\star,\pm}_i $ stands for the left(right) $\pm$ polarized component
of the incoming or detected mode electric field amplitude.
The dipole moments of transitions (in units of $\tilde{W}_i /(\mathcal{E}^{\pm}_i \sqrt{N_i})$) are:
\begin{eqnarray}
\label{EQ:2_20}
\mu^{\pm}_{m1,m2} =
\int d \mathbf{r} \langle{\Psi_{m1}(\mathbf{r})}\vert r \texttt{e}^{\pm i \theta} \vert{\Psi_{m2}(\mathbf{r})}\rangle
\end{eqnarray}•
Note that we are still in the single electron-hole representation, not yet in the many body exciton/hole
representation.
Therefore we do not need the envelop function to define the transition moments as in
Ref.~\cite{agranovich1992electrodynamics}.

\par
The next step is to bring the Hamiltonian in Eq. \eqref{EQ:2_18},\eqref{EQ:2_19}
into the excitonic form. Using the method first proposed in Ref.\cite{chernyak1998multidimensional}  
we define the electron-hole pair annihilation operators (not to be confused
with exciton operators) as:

\begin{gather}
\label{EQ:2_21}
\hat{B}^{\dag}_m = \hat{B}_{m1,m2} = \hat{c}_{m1} d_{m2}
\end{gather}•
where we used composite index of
$m = \left\{{m1,m2}\right\}$
Since we are interested in the third-order response the commutator of the above operators may
be truncated at quadratic order:
\begin{equation}
\left[{\hat{B}_{m},\hat{B}^\dag_n}\right] = \delta_{m,n} -
2 \sum_{p,q} \delta_{m,n;p,q} \hat{B}^\dag_p \hat{B}_q
\end{equation}•
where $\delta_{m,n} = \delta_{m1,n1} \delta_{m2,n2}$.
The tetradic matrix $\delta_{m,n;p,q}$ (phase-filling factor) is
responsible for the deviation from the boson statistic of the
pair operators, and steams from the fermionic nature of its constituents:

\begin{gather}
\label{EQ:2_22}
2 \delta_{m,n;p,q}  \\
\notag
=\delta_{m1,q1}
\delta_{m2,p2}
\delta_{n1,p1}
\delta_{n2,q2}
+
\delta_{m1,p1}
\delta_{m2,q2}
\delta_{n1,q1}
\delta_{n2,p2}
\end{gather}•
In the basis of electron-hole pairs the Hamiltonian in Eqs. \ \eqref{EQ:2_17}, \eqref{EQ:2_18} becomes
Frenkel-like if truncated up to forth order(valid for third order response with two excited electron-hole pairs):
\begin{gather}
\label{EQ:2_23}
\hat{\mathcal{H}}  
= \sum \limits_{m} E_{m} \hat{B}^\dag_m \hat{B}_m + \frac{1}{4}
\sum \limits_{m,n} \left({E_m + E_n}\right) \hat{B}^\dag_m \hat{B}^\dag_n \hat{B}_m \hat{B}_n\\
\hat{\mathcal{H}}_{\texttt{int}}(t)   
= \sum \limits_{i,m}
\sum \limits_{j =\pm}
\mathcal{E}^j_i \delta(t-t_i) \mu^j_{m} \hat{B}_m +\texttt{H.c}
\end{gather}•
Direct diagonalization of the above Hamiltonian \eqref{EQ:2_23}
in order to find the exciton/biexciton manifolds is difficult and non-equilibrium Green's functions for the single and double electron-hole pairs are used instead.
If one neglects the nonlinearities caused by the Pauli exclusion those retarded Green's
functions are defined as:
\begin{gather}
\label{EQ:2_24}
G_{e1} (\tau) = - i \theta(\tau) \texttt{e}^{-i E_{e1} \tau}\\
\label{EQ:2_25}
G_{e1,e2} = -i \theta(\tau) \texttt{e}^{-i (E_{e1} + E_{e2}) \tau}
\end{gather}
Here $\theta(\tau)$ is the Heaviside function
and the time between two consecutive pulses is denoted as $\tau_{i} = t_{i+1} - t_i$.
Note that in order to be retarded the
Green's functions must contain the energies with $\texttt{Im}[E_e]<0$.
We also adopted the notation $E_{e1} = E_{m1}+E_{m2}$.
We shall also need their Fourier transforms with respect to the time delays:
\begin{gather}
\label{EQ:2_26}
G_{e1}(\omega) = \frac{1}{\omega - E_{e1}}\\
\label{EQ:2_27}
G_{e2,e1}(\omega) = \frac{1}{\omega - E_{e1}-E_{e2}}
\end{gather}
In the above Green's functions the
biexciton energies (the poles of Eq. \eqref{EQ:2_27}) is simply a sum of the exciton energies.
The nonlinear signal from such system vanishes since it represents a collection of harmonic oscillators.
The effect of Pauli exclusion in Eq. \eqref{EQ:2_23} is usually incorporated by tetradic exciton scattering matrix:
\begin{gather}
\label{EQ:2_28}
\Gamma_{e4,e3;e2,e1}(\omega) =
\delta_{e4,e3;e2,e1} (G^{-1})_{e4,e3}
\end{gather}
Which carries all the information about underlying nonlinearities.
Coulomb interaction can be incorporated
by solving Bethe-Saltpeter equation as
in Ref.\cite{mukamel2007sum,chernyak1998multidimensional,mukamel1995principles}.

\par
The photon echo signal can be recast in terms of non-interacting Green's functions as well as
the scattering matrix as:
\begin{gather}
\label{EQ:2_29}
S^{j1,j2,j3,j4}_{-\mathbf{k}_1 +\mathbf{k}_2 + \mathbf{k}_3}
\left({\omega_3},\tau_2 =0,\omega_1 \right) \\
\notag
= 2 \texttt{Re} \sum \limits_{e1,e2,e3,e4}
\mu^{j1}_{e3} \mu^{\star,j2}_{e2} \mu^{\star,j3}_{e1} \mu^{j4}_{e4}
G^\star_{e3}(-\omega_1) G_{e4}(\omega_3) \\
\notag
\times \Gamma_{e4,e3;e2,e1} (\omega_3 + E_{e3})
G_{e2,e1}(\omega_3 + E_{e3})
\end{gather}

\begin{figure}[]
\centering
\includegraphics[width=0.8\columnwidth]{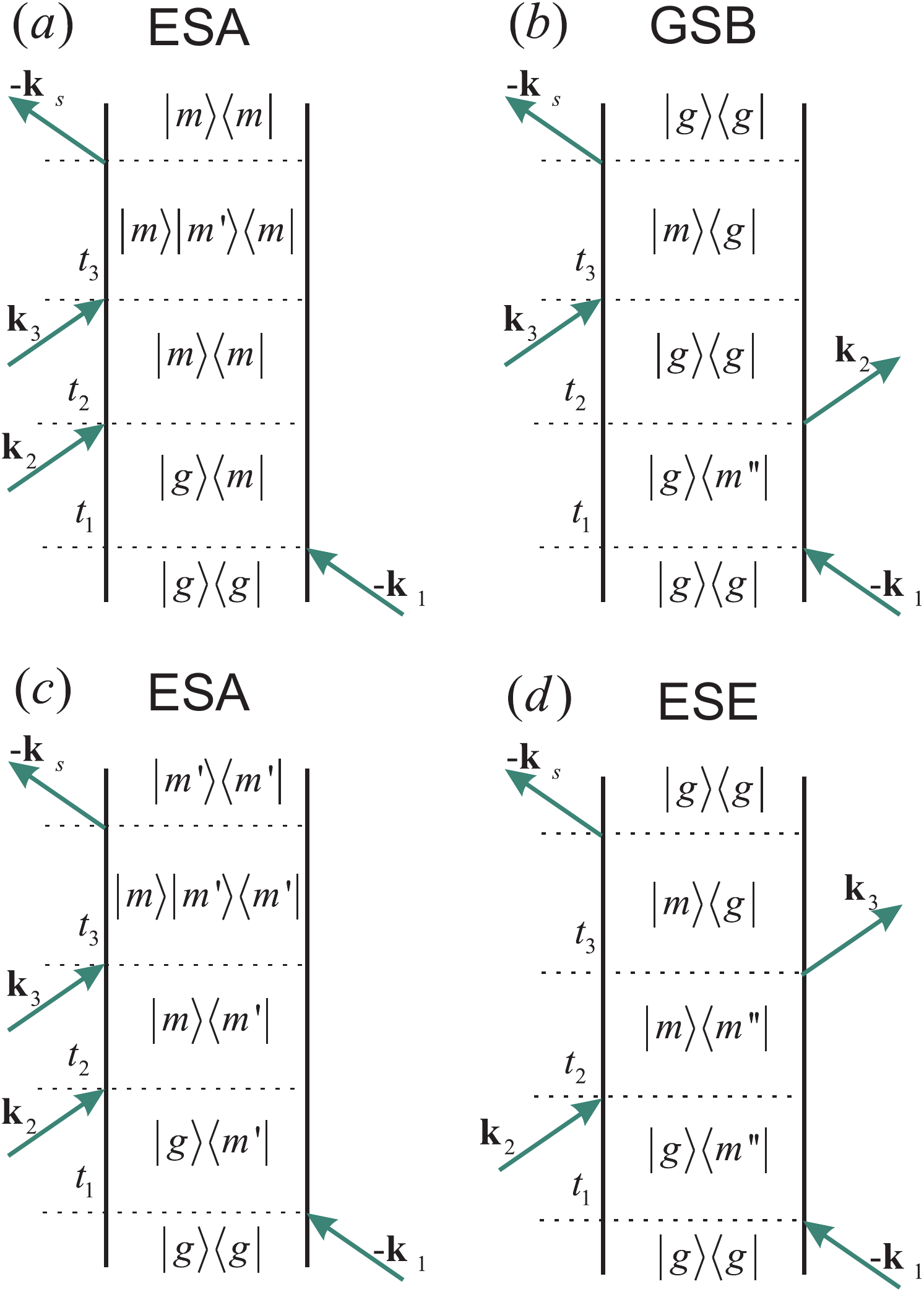}
\caption{\label{FIG:5} Feynman diagrams for the photon echo technique $\mathbf{k}_s = -\mathbf{k}_1 + \mathbf{k}_2 + \mathbf{k}_3$.}
\end{figure}

Even though the above two expression formally gives the nonlinear signal, it is hard to analyze.
However it is convenient for numerical simulations due to expandability of the scattering matrix into the domain
where the coulomb interaction may play its role. 
The detailed form of the scattering matrix which involves the Coulomb interaction is given in Ref.\cite{kavousanaki2009probing,mukamel2007sum,chernyak1998multidimensional}.
An alternative approach to derive the signal by using double-sided time ordered Keldish diagrams shown in Fig. \ref{FIG:5}.
The diagrammatic approach (also known as "sum over states") can answer
one of the fundamental question whether the nonzero scattering matrix is sufficient to calculate a 
nonlinear signal. 
The answer to that question is not trivial due to
large number of interfering terms in Eqs. \eqref{EQ:2_29} and \eqref{EQ:2_30}.
The diagrams were constructed by blocking the consequent double
excitation of the same electron-hole pair. 
The nonlinear signals can be extracted from the
diagrams by the rules stated in Ref.\cite{mukamel2007sum,abramavicius2009coherent}.
In our case the photon echo signal is obtained via the diagrams in Fig. \ref{FIG:5}:
\begin{gather}
\label{EQ:2_32}
S^{j1,j2,j3,j4}_{-\mathbf{k}_1 +\mathbf{k}_2 + \mathbf{k}_3}
\left({\omega_3},\tau_2 ,-\omega_1 \right) \\
\notag
= \texttt{Re} \sum \limits_{m, m' \neq m}
\frac{\mu_m^{j1}}{-\omega_1 - E_m} \frac{\mu^{\star,j2}_{m} \mu^{\star,j3}_{m'} \mu^{j4}_m  \texttt{e }^{-i \tau_2  (E_m - E^\star_{m})}}{\omega_3 - E_m - E_m' + E_m} \\
\notag
+ \frac{\mu_{m'}^{j1}}{-\omega_1 - E_{m'}} \frac{\mu^{\star,j2}_{m'} \mu^{\star,j3}_{m} \mu^{j4}_{m'} \texttt{e }^{- i \tau_2  (E_m - E^\star_{m'})}}{\omega_3 - E_m - E_m' + E_m'} \\
\notag
- \texttt{Re} \sum \limits_{m, m''}  \frac{\mu_{m''}^{j1}}{-\omega_1 + E_{m''}}
[\frac{\mu^{\star,j2}_{m} \mu^{\star,j3}_{m''} \mu^{j4}_m}{\omega_3 - E_m} \\
\notag
+ \frac{\mu^{\star,j2}_{m''} \mu^{\star,j3}_{m} \mu^{j4}_{m''} \texttt{e}^{-i \tau_2 \left ({E_m - E^\star_{m'}}\right)} }{\omega_3 - E_{m}}] \ .
\end{gather}
This signal would vanish if it were not for the Pauli blocking which prevents $m$ to be equal to $m'$.
\par
In Appendix \ref{AP:4} we have analyzed an alternative form of the four-wave mixing known as double-quantum coherence.
This signal vanishes identically despite the Pauli induced scattering since the exited state absorption pathways are fully compensated by
their ground state bleaching and exited state emission counterparts. 
Therefore the double-quantum-coherence can be readily used as a measure of the Coulomb interaction strength, and screening.

\section{\label{SEC:6} Numerical results and discussion}

The main advantage of the Pauli blocking description of excitons is its simplicity.
For a model with $N$ singly excited electronic states, we only need to consider $N$ double
excited states compared with $N(N-1)$ in the case of Coulomb scattering induced biexcitons.
We note that the same number of doubly excited states $N(N-1)$ are allowed in a simple boson
harmonic model. 
This allows for better tracking of pathways interference
and resonances. 
In this section, we shall use it to classify the off-diagonal
resonances in the 2D photon echo spectra in accordance with the short living excited
states of the QD. This will be compared with the linear absorption spectrum
which is proportional to the single excited electronic density of states and is given
by the main diagonal of the 2D spectra. 
We shall demonstrate the improved resolution of those
short living states via the coherent response with the long living excitation.
First, we assume the a model of two single excited states $(\mu-1,E_1;\ \mu_2, E_2)$.
This leads by the Pauli exclusion principle to the single double excited state
$(\mu_{12},E_1 + E_2)$. 
The photon echo signal contains three distinct pathways:
ground state bleaching (GSB Fig. \ref{FIG:5}(b)), excited
state emission (ESE Fig. \ref{FIG:5}(d), and excited state absorption (ESA Fig. \ref{FIG:5}(a,c).
Those are given by

\begin{gather}
\label{EQ:3_1}
S^{(3)}_{\texttt{GSB}}(\omega_3,t_2,\omega_1) =
\texttt{Re} \sum \limits_{i ,j=1}^{2} \frac{\vert{\mu_i}
\vert^2 \vert{\mu_j}\vert^2}{\left({-\omega_1 -E^\star_i}\right)
\left({\omega_3 -E_j}\right)}\\
\label{EQ:3_2}
S^{(3)}_{\texttt{ESE}}(\omega_3,\tau_2,\omega_1) =
\texttt{Re} \sum \limits_{i ,j=1}^{2} \frac{\vert{\mu_i}\vert^2 \vert{\mu_j}\vert^2 \texttt{e}^{-\tau_2(E_j-E^\star_i)}}{\left({-\omega_1 -E^\star_i}\right) \left({\omega_3 - E_j}\right)}\\
\label{EQ:3_3}
S^{(3)}_{\texttt{ESA}}(\omega_3,\tau_2,\omega_1)  \\
\notag
=-  \texttt{Re} \sum \limits_{i \neq j}^{2} \frac{\vert{\mu_i}\vert^2 \vert{\mu_j}\vert^2 \texttt{e}^{-\tau_2(E_i-E^\star_i)}}{\left({-\omega_1 -E^\star_i}\right) \left({\omega_3 - E_j-E_i + E^\star_i}\right)} \\
\notag
-  \texttt{Re} \sum \limits_{i \neq j}^{2} \frac{\vert{\mu_i}\vert^2 \vert{\mu_j}\vert^2 \texttt{e}^{-\tau_2(E_i-E^\star_j)}}{\left({-\omega_1 -E^\star_j}\right) \left({\omega_3 - E_i-E_j+E^\star_j}\right)} \ .
\end{gather}
Clearly,   when Pauli blocking is neglected, we have a collection of
damped oscillators and at $\tau_2 = 0$ the signal disappears. 
It would also vanish
if we assume that there is  no damping in the system. 
We note that Pauli blocking may be suppressed
 when the double exciting state is formed by electron/hole
pairs with opposite spins.

\par
\begin{figure}[]
\centering
\includegraphics[width=0.8\columnwidth]{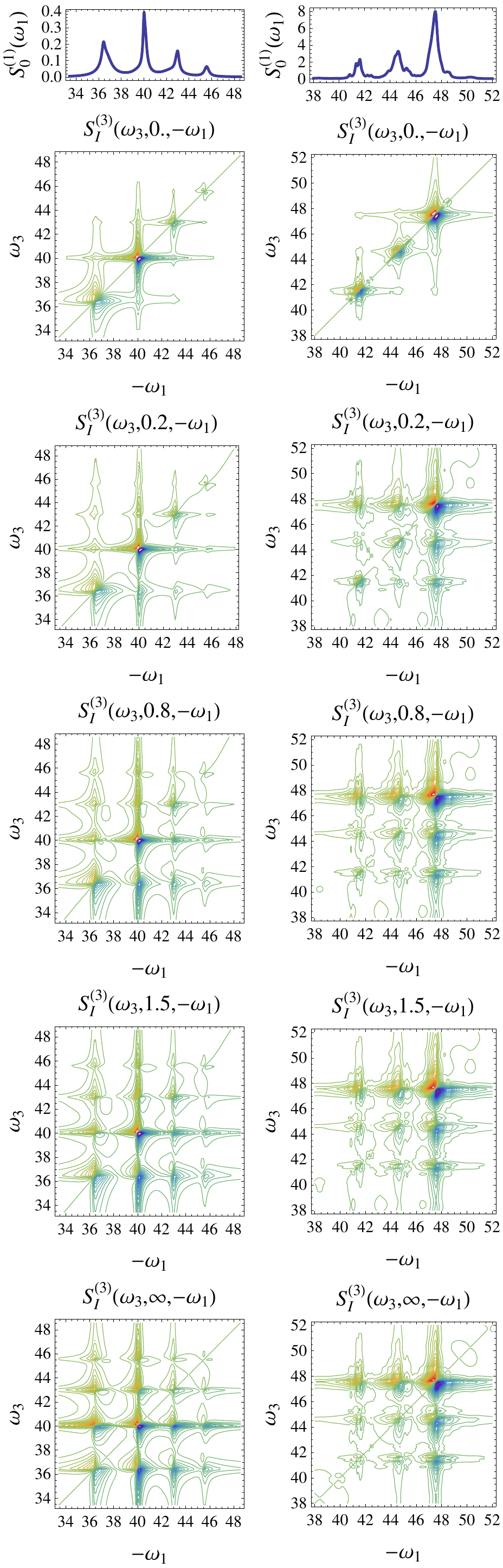}
\caption{\label{FIG:6}Photon echo signal for various time delays $\tau_2$ calculated 
from Eq.~\eqref{EQ:2_32}. 
The left panels correspond to $E_g =0$, while the
 right panels demonstrate the dressing effect $E_g =10$. The top-marginal
graphs show the linear absorption.}
\end{figure}

Since the nonlinear signal vanishes for ideal bosons, one can recast it
to the alternative simplified form as if from the ESA from otherwise Pauli
blocked $N$ states as

\begin{gather}
\label{EQ:3_4}
S^{(3)}_{-\mathbf{k}_1 +\mathbf{k}_2 + \mathbf{k}_3} \left({\omega_3,0 ,-\omega_1 }\right)
 \\
\notag
= 2 \texttt{Re} \sum \limits_{i=1}^{N} \frac{\vert{\mu_i}\vert^2
\vert{\mu_j}\vert^2}{\left({-\omega_1 -E^\star_i}\right) \left({\omega_3 -E_i-E_j + E^\star_j}\right)}
\end{gather}
At zero time delay $\tau_2 =0$ we have only the diagonal resonances.
The states with small damping dominate the picture.
However as the time delay progresses the off diagonal resonances appear, as
demonstrated in Fig. \ref{FIG:6}.
It is convenient to interpret the signal by comparing it with the linear absorption
(the top-marginal graph in the figure). 
For our numerical simulations, we chose the potential
kink height to be $V_1=-V_2 = 20$. 
Unless stated otherwise, all the energies are in units
of $3 \tilde{\gamma}_0 a/R$. We have also added constant dephasing rate $\gamma = 0.1$ to
account for possible contact with an external phonon bath.  
We have also limited ourselves to
electronic states with angular momentum up to $m=9/2$.
An idealized single graphene layer grown epitaxially on $\texttt{SiO}$ was considered.

\par
The tallest absorption peak comes from the transition $5/2 \to 7/2$,
and is located at $\omega_1 \approx 40.0$. 
That signifies the true bound state when the
electron (hole) energy reaches the height of the potential barrier (See Eq. ~\eqref{EQ:2_15} and
Fig. \ref{FIG:3}). 
The remaining absorption peaks correspond to quasi-bound states
with finite lifetime. 
For the linear spectrum, the latter brings the peak broadening.
To extract additional information about the dynamics of the quasi-bound states, we resort to
the photon echo signal.
At zero time delay $\tau_2 = 0$,
this provides the same information as the linear absorption. 
The positions and magnitudes of the cross-peaks (rapid change in the sign of
the signal) on the main diagonal correspond to those in the linear absorption.
The existence of the signal comes from the fact that the electrons (holes) are not
coupled to a simple bath of harmonic oscillators (constant dephasing). 
The pattern
of the cross-resonances along the main diagonal is the manifestation of the destructive
interference between the GSB, ESE on one side and the ESA pathways on the other.
The latter takes into account the Pauli blocking effect on the biexciton
(two electron-hole pair) states. 
At this point, we completely neglected the effect due
to  the Coulomb interaction between electrons.  Later, we shall demonstrate that
it is a reasonable assumption for small QDs.  
Thanks to the very simple exciton scattering
matrix based on Pauli blocking, only (Eq.~\eqref{EQ:2_28}), we can employ a
 simplified quasi-particle picture in order to describe the signal (Eq.~\eqref{EQ:3_4}).

\par
By increasing the time delay
$\tau_2$, we may monitor the lifetime dynamics of the quasi-bound  states as follows.
The ESA and ESE contributions to the signal is reduced and  finally the
GSB signal survives (the lowest graphs in Fig. \ref{FIG:6}).
In between, the off-diagonal cross-peaks  appear  at a time.
Those with the smaller dephasing rate (strongly bound to the QD) appear first.
The most pronounced cross-peaks are those which are correlated to the true bound state.

\par
We next turn our attention to the dressed Dirac electrons confined to the
potential   induced QD.  
The dressing opens up a dynamical gap
which can be controlled by the intensity and polarization degree of CW pumping light.
We shall probe the dynamic gap by the photon echo technique described above and
compare it to the linear absorption. 
The gap allows for many more bound states
since the wave vector of the outgoing electronic wave $\kappa$ can cross over into the
imaginary plane, thus effectively quenching the outgoing wave and bounding the electron
states (See Fig. \ref{FIG:3}). 
For our simulations, we chose the   gap
$E_g = 10$ which may be achieved either for small QD or intense   pumping field
with circular polarization. We note that the gap may also be induced by a
 polar substrate\cite{qaiumzadeh2009ground}.
The gap achieves several bound states for the
 $1/2 \to  3/2$ electronic transitions. Since the wave functions for
 larger angular momentum are highly oscillatory, the latter transition posses highest
 oscillator strength, thereby effectively shifting the position of the main peak
in the linear absorption (see Fig. \ref{FIG:6} right panel).
The remainder of the peaks also contain a mixture between the bound and
quasi-bound states. 
To separate these, we shall look at the photon echo at $\tau_2  >0$.
Finally, the resulting GSB  reveals the truly bound states (see Fig. \ref{FIG:6} right panel).

\par
To examine the role played by   Coulomb scattering, we shall employ the full form of
the scattering matrix. 
The approach is based on the nonlinear exciton equations (NEE).
We refer the reader to the comprehensive review of the technique given by
Abramavicus and   Mukamel \cite{abramavicius2009coherent}.
Exciton scattering is best described in the eigenstate basis of Eqs.~\eqref{EQ:2_14} and
\eqref{EQ:2_13}. Keeping in mind that we can have at most two excitons, leads to
effective truncation schematics of otherwise
infinite series of intertwined NEEs\cite{mukamel1995principles}. 
In the latter case, an appropriate factorization scheme has been applied.
We have also neglected  incoherent exciton transport.

\begin{figure}[]
\centering
\includegraphics[width=0.8\columnwidth]{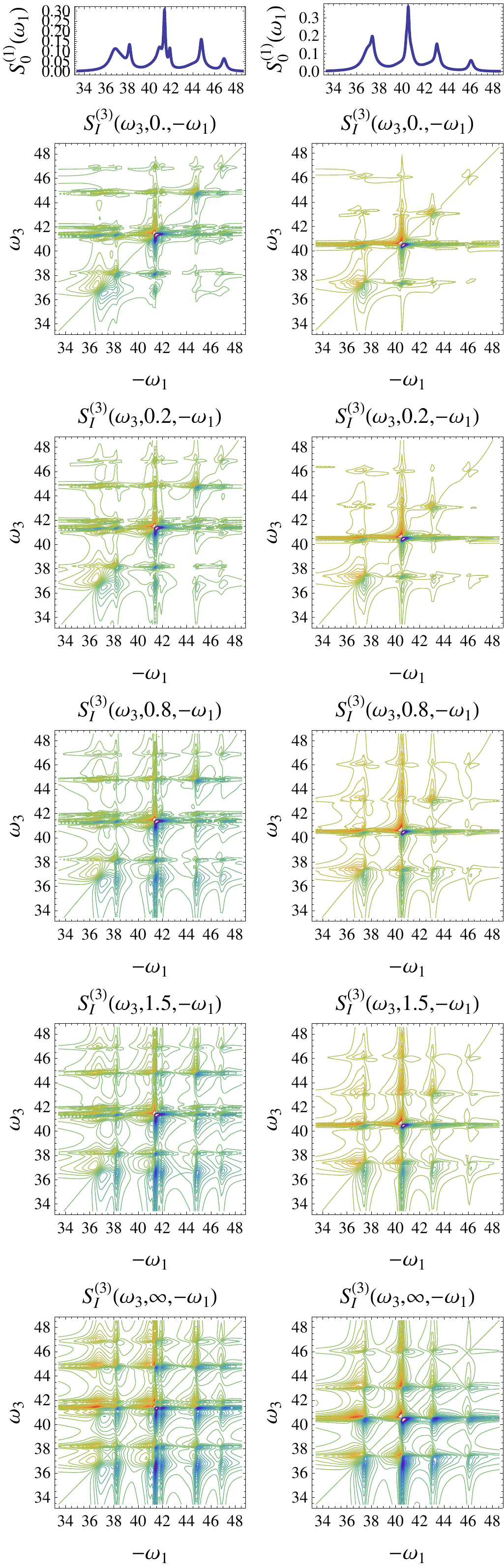}
\caption{\label{FIG:7}The same as in Fig. \ref{FIG:6} with the Coulomb scattering taken
into account via Eq.~\eqref{EQ:2_29} with the scattering matrix given by Eq.
The left panels represent large QD, $R/a_0 = 1000$. Right panels
are for smaller QD, $R/a_0 = 100$.}
\end{figure}

\par
The photon echo and the linear absorption are shown in the right
panel of Fig. \ref{FIG:7} for larger size QD. 
We see the off-diagonal
correlation resonances and symmetry breaking for $\tau_2 =0$. These indicates the
bonding and anti-bonding biexciton resonances with the biexciton binding energy of
a few $\texttt{eV}$. Indeed, when the biexciton binding energy is increased as a
result of the Coulomb interaction, the ESA peaks are shifted along $\omega_3$:
downwards for positive anti-binding (exciton repulsion) and upwards for negative
bonding energy (exciton attraction). The ESA cross-peaks are no longer cancelled by the
GSB and ESE, thus creating the doublets. By increasing the QD size, we see the
formation of excitons with  exciton binding energy of $10-30 \; \texttt{eV} $
in the left panel of Fig. \ref{FIG:7}. Signatures of the off-diagonal quadratic
coupling also persist for longer delay times $\tau_2>0$.

\section{\label{SEC:7} Concluding Remarks}

We proposed dressing the Dirac electrons with   circularly polarized photons
in order to localize them within a QD on graphene monolayer. We also investigated
the localization of dressed electrons in a cylindrical QD formed on  bilayer graphene.
When graphene is irradiated with a circularly polarized electromagnetic field,
An energy gap opens up in the dispersion relation for graphene in the presence of
this electromagnetic field.  Consequently,
the resulting confined electronic states for a QD seem to have  properties
that are similar in nature to the surface states
of topological insulators. Their energies are located inside the energy gap and the
wave functions  decay as a function of distance from the interface of the  potential.
These topological
states are robust with respect to the effects of disorder.
Our calculations showed that the dressing does not only open
a dynamical gap in the energy dispersion spectrum,  but it also leads to a renormalization
of the Fermi velocity as well as the intra layer and
inter layer coupling parameters. In fact, in the bilayer configuration,
the dressing serves as a tool for  tuning  the energy gap. That is, it can either
 close or open the gap, depending on
the polarity of the potential  and the direction of the light polarization.
 Linear spectroscopy cannot resolve the short lived broadened excitonic states
 and must be resolved  by using a four-wave mixing technique  known as
photon-echo. 
This eliminates the inhomogeneous
broadening due to impurities, and  to focuses on the intrinsic lifetimes of
the electronic states.
We measure  the localization through the electronic density of states,
The strong dynamical screening of the Coulomb interaction  leads  us to consider
only the Pauli blocking due to  the Fermi statistics.   We simplify the signal
 interpretation by switching to the quasiparticle picture. Those are give as the
 deviation from the harmonic oscillator for which the nonlinear signals disappear.
This allows us to consider only excited states absorption Liouville pathways.
In this way, we are able to reduce the interference due to
the usual combination  between the ground state bleaching and excited
states emission. Visible light is used to map the QD interband
transitions onto 2D spectra and terahertz pulse shaped fields for the intraband
transitions.
Important aspects of terahertz pulse shaped fields for the intraband
transitions will be reported elsewhere. 
The latter will allow us to use a novel and more convenient phase cycling
method to obtain the $\chi^{(3)}$ response \cite{kuehn2011strong}.

\begin{acknowledgments}
The authors
gratefully acknowledge the support of
Air Force Research Lab (AFRL) by  contract No.\#    FA 9453-11-01-0263; 
the National Science Foundation (NSF) through Grant No.\# CHE-1058791, DARPA BAA-10-40 QuBE;
from Chemical Sciences, Geosciences, and Biosciences
Division, Office of Basic Energy Sciences, Office of Science,
(U.S.) Department of Energy (DOE).
\end{acknowledgments}

\bibliography{biblio}

\begin{appendix}
\section{\label{AP:1}Derivation of Eq.~\eqref{EQ:1_1}}

The original Hamiltonian has the form

\begin{equation}
\label{EQA:1_1}
\mathcal{H} = \hbar v_F \boldsymbol \sigma \cdot \left({\mathbf{k}-\frac{e}{c \hbar}\mathbf{A}}\right)
\end{equation}
The vector potential operator of the electromagnetic field
can be partitioned as:
\begin{gather}
\label{EQA:1_2}
\mathbf{A} = \mathbf{A}_0 + \sum \limits_{i = 1}^\infty \mathbf{A}_i\\
\label{EQA:1_3}
\mathbf{A}_0 = \sqrt{\frac{ 2 \pi \hbar c^2}{\omega_0 \Omega}}
\left({\mathbf{e}_{+} a_0 + \mathbf{e}_- a^\dag_0}\right)\\
\label{EQA:1_4}
\mathbf{A}_i = \sqrt{\frac{ 2 \pi \hbar c^2}{\omega_i \Omega}}
\left({\mathbf{e}_{x} (a_i + a^\dag_i)}\right)
\end{gather}
Here $\sqrt{2} \mathbf{e}_{\pm} = \mathbf{e}_x \pm i \mathbf{e}_y$
are polarization vectors given in terms of the unit vectors along corresponding Cartesian directions;
$\Omega$ is the mode quantization volume.
As one can see Eq.~\eqref{EQA:1_2} describes the electromagnetic wave propagating along $z-$axis (transverse to graphene).
It is clock-wise circularly polarized.
We will need the circular polarization since graphene is gapless and no RWA is applicable.
The rest of the optical modes described by Eq.~\eqref{EQA:1_3} are linearly polarized.
Note that we have no phase on the optical filed since we assume graphene being ideally
flat and situated at $z=0$. That is $\exp (\pm i k_z z) =1$.
Substituting Eq.~\eqref{EQA:1_2} into Eq.~\eqref{EQA:1_1}
and denoting $-e \sqrt{4 \pi \hbar v^2_F/\omega_i \Omega} = W_i/\sqrt{N_i}$ in order to
keep notation consistent with Ref.~\cite{gerry2003introduction} we obtain Eq.~\eqref{EQ:1_1}.

\section{\label{AP:2}Derivation of Eqs.~\eqref{EQ:1_10} and \eqref{EQ:1_11}}

We first need the following identities

\begin{gather}
\label{EQA:2_1}
\sigma_{\pm} \vert{\pm, N_0} \rangle =  \vert{\mp, N_0} \rangle\\
\label{EQA:2_2}
\sigma_{\pm} \vert{\mp, N_0} \rangle = 0
\end{gather}
Therefore, we shall have:
\begin{gather}
\label{EQA:2_3}
\hbar v_F \left({\sigma_x k_x + \sigma_y k_y}\right) \vert{ \psi_{\pm,N_0}} \rangle  \\
\notag
=\hbar v_F \left({(\sigma_+ + \sigma_-) k_x + i (\sigma_- - \sigma_+) k_y }\right)
\vert{\psi_{\pm,N_0}} \rangle \\
\notag
=\hbar v_F \left({(\sigma_+ + \sigma_-) k_x + i (\sigma_- - \sigma_+) k_y }\right)  \\
\notag
\times \left({
\cos \phi \vert{\pm,N_0}\rangle \pm
\sin \phi \vert{\mp,N_0 \pm 1}\rangle}\right) \\
\notag
=\hbar v_F (
k_x \left({
\cos \phi \vert{\mp, N_0}\rangle \pm \sin \phi \vert{\pm, N_0 \pm 1}\rangle
}\right)
\pm \\
\notag
\pm
i k_y \left({
\cos \phi \vert{\mp, N_0}\rangle \mp \sin \phi \vert{\pm, N_0 \pm 1}\rangle
}\right)
)
\end{gather}

Using the above equation we can calculate all the necessary matrix elements:
\begin{gather}
\notag
\mathcal{H}_1 = \tilde{\mathcal{H}}_1 + V(x,y)\\
\label{EQA:2_4}
\langle{\psi_{+, N_0}}\vert \tilde{\mathcal{H}}_1 \vert{\psi_{+, N_0}}\rangle  \\
\notag
= \hbar v_F (
(\cos \phi \pl{N_0} + \sin \phi \ml{N_0 +1})\\
\notag
\times k_x \left({
\cos \phi \vert{-, N_0}\rangle + \sin \phi \vert{+, N_0 + 1}\rangle
}\right)
  \\
\notag
+ i k_y \left({
\cos \phi \vert{-, N_0}\rangle - \sin \phi \vert{+, N_0 + 1}\rangle
}\right)
) = 0\\
\label{EQA:2_5}
\langle{\psi_{-, N_0}}\vert \tilde{\mathcal{H}}_1 \vert{\psi_{-, N_0}}\rangle  \\
\notag
=\hbar v_F (
(\cos \phi \ml{N_0} - \sin \phi \pl{N_0 -1})\\
\notag
\times k_x \left({
\cos \phi \vert{+, N_0}\rangle - \sin \phi \vert{-, N_0 - 1}\rangle
}\right)
  \\
\notag
-
i k_y \left({
\cos \phi \vert{+, N_0}\rangle + \sin \phi \vert{-, N_0 - 1}\rangle
}\right)
) = 0\\
\label{EQA:2_6}\\
\notag
\langle{\psi_{-, N_0}}\vert \tilde{\mathcal{H}}_1 \vert{\psi_{+, N_0}}\rangle  \\
\notag
=\hbar v_F (
(\cos \phi \ml{N_0} - \sin \phi \pl{N_0 -1})\\
\notag
\times k_x \left({
\cos \phi \vert{-, N_0}\rangle + \sin \phi \vert{+, N_0 + 1}\rangle
}\right)
  \\
\notag
+  i k_y \left({
\cos \phi \vert{-, N_0}\rangle - \sin \phi \vert{+, N_0 + 1}\rangle
}\right) )   \\
\notag
= \cos^2 \phi \left({k_x + i k_y}\right)\\
\label{EQA:2_7}
\langle{\psi_{+, N_0}}\vert \tilde{\mathcal{H}}_1 \vert{\psi_{-, N_0}}\rangle  \\
\notag
=\hbar v_F (
(\cos \phi \pl{N_0} + \sin \phi \ml{N_0 + 1})\\
\notag
\times k_x \left({
\cos \phi \vert{+, N_0}\rangle - \sin \phi \vert{-, N_0 - 1}\rangle
}\right)
  \\
\notag
-  i k_y \left({
\cos \phi \vert{+, N_0}\rangle + \sin \phi \vert{-, N_0 - 1}\rangle
}\right) )   \\
\notag
= \cos^2 \phi \left({k_x - i k_y}\right)\\
\label{EQA:2_7}
\langle{\psi_{\pm, N_0}}\vert V(x,y) \vert{\psi_{\pm, N_0}}\rangle  \\
\notag
= V(x,y) \left({\cos^2 \phi + \sin^2 \phi}\right) = V(x,y)
\end{gather}

For $\mathcal{H}_2$ matrix elements we will need, the following identities.
\begin{gather}
\label{EQA:2_8}
\langle{\psi_{\pm, N_0}}\vert \sigma_{+} + \sigma_{-} \vert{\psi_{\pm, N_0}}\rangle \\
\notag
= \langle{\psi_{\pm, N_0}} \vert
\left({\cos \phi \vert{\mp, N_0}\rangle \pm \sin
\phi \vert{\pm, N_0 \pm 1}\rangle}\right) \\
\notag
= \left({
\cos \phi \langle{\pm, N_0}\vert \pm \sin \phi \langle{\mp, N_0 \pm 1}\vert
}\right)   \\
\notag
\times \left({\cos \phi \vert{\mp, N_0}\rangle
\pm \sin \phi \vert{\pm, N_0 \pm 1}\rangle}\right) =0\\
\label{EQA:2_9}
\langle{\psi_{\mp, N_0}}\vert \sigma_{+} + \sigma_{-} \vert{\psi_{\pm, N_0}}\rangle
= \cos^2 \phi
\end{gather}

\section{\label{AP:3} Derivation of Eqs.~\eqref{EQ:2_5} and \eqref{EQ:2_6}}

For the b-layer we will need the following identities:
\begin{gather}
\label{EQ:1_9}
\langle{\psi_{N_0}}\vert \sigma_{\pm} \vert{\psi_{N_0}}\rangle   \\
\notag =\left({
\begin{array}{cc}
\langle{\psi_{+,N_0}}\vert \sigma_{\pm} \vert{\psi_{+,N_0}}\rangle
&
\langle{\psi_{+,N_0}}\vert \sigma_{\pm} \vert{\psi_{-,N_0}}\rangle\\
\langle{\psi_{-,N_0}}\vert \sigma_{\pm} \vert{\psi_{+,N_0}}\rangle
&
\langle{\psi_{-,N_0}}\vert \sigma_{\pm} \vert{\psi_{-,N_0}}\rangle
\end{array}
}\right) \\
\notag
=\left({\cos^2 \phi}\right) \sigma_{\pm}
\end{gather}

\begin{gather}
\label{EQ:1_9}
\langle{\psi_{N_0}}\vert \sigma_{+} a _0 \vert{\psi_{N_0}}\rangle =
\langle{\psi_{N_0}}\vert \sigma_{-} a^\dag _0 \vert{\psi_{N_0}}\rangle  \\
\notag
= \frac{\sqrt{N_0}}{2} \sin 2 \phi
\left({
\begin{array}{cc}
1
&
0\\
0
&
-1
\end{array}
}\right)
= \left({\frac{\sqrt{N_0}}{2} \sin 2 \phi }\right) \sigma_3
\end{gather}

\section{\label{AP:4} Double quantum coherence.}
\begin{figure}[]
\centering
\includegraphics[width=0.8\columnwidth]{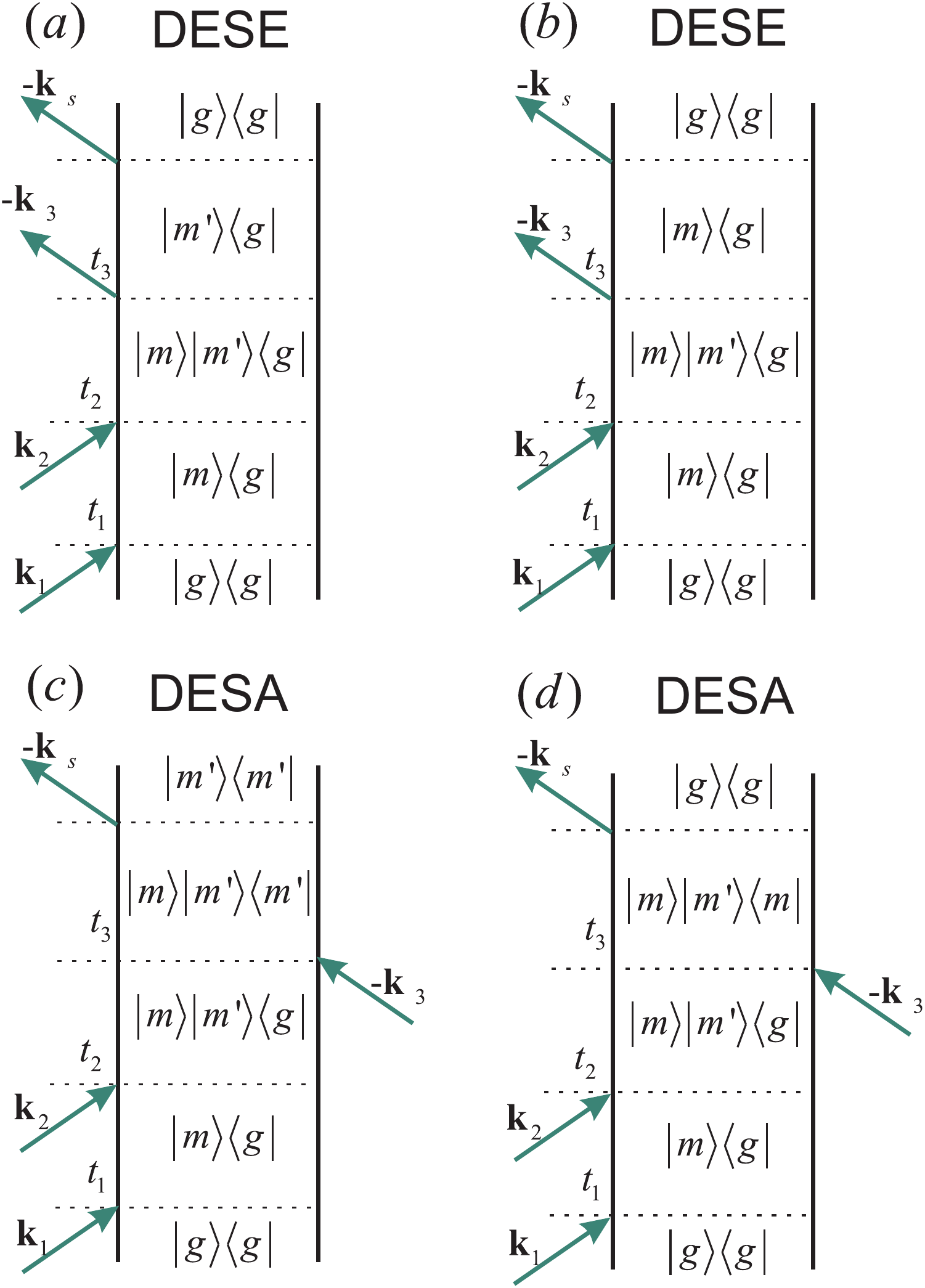}
\caption{\label{FIG:4} Feynman diagrams for the double quantum coherence signal generated in $\mathbf{k}_s =\mathbf{k}_1 + \mathbf{k}_2 - \mathbf{k}_3$ direction.}
\end{figure}

The double quantum coherence signal can be derived from the diagrams in Fig.~\ref{FIG:4} assuming the following form:
\begin{gather}
\label{EQ:2_30}
S^{j1,j2,j3,j4}_{\mathbf{k}_1 +\mathbf{k}_2 - \mathbf{k}_3}
\left({\omega_3},\omega_2,\tau_1 =0 \right) \\
\notag
= 2 \texttt{Im} \sum \limits_{e1,e2,e3,e4}
\mu^{\star,j1}_{e1} \mu^{\star,j2}_{e2} \mu^{j3}_{e3} \mu^{j4}_{e4}
G_{e4}(\omega_3) G^\star_{e3}(\omega_2 - \omega_3) \\
\notag
\times \left[{
\Gamma_{e4,e3;e2,e1} (\omega_3 + E_{e3})
G_{e2,e1}(\omega_3 + E_{e1})- }\right.\\
\notag
\left.{
-
\Gamma_{e4,e3;e2,e1}(\omega_2) G_{e2,e1}(\omega_2)
}\right]
\end{gather}

When the Coulomb scattering may be neglected the above signal is greatly simplified into sum-over-states expression with
the explicit Pauli blocking principle:
\begin{gather}
\label{EQ:2_31}
S^{j1,j2,j3,j4}_{\mathbf{k}_1 +\mathbf{k}_2 - \mathbf{k}_3} \left({\omega_3},\omega_2,\tau_1 \right) \\
\notag
=  \texttt{Re} \sum \limits_{m, m' \neq m}
\frac{\mu^{\star,j1}_{m} \mu^{\star,j2}_{m'} \texttt{e}^{-i E_m \tau_1} }{\omega_2 - E_m - E_{m'}} \\
\notag
\times \left[{ \frac{\mu^{j3}_{m} \mu^{j4}_{m'}}{\omega_3 - E_m - E_{m'} + E_{m'}}+
\frac{\mu^{j3}_{m'} \mu^{j4}_{m}}{\omega_3 - E_m - E_{m'} + E_{m}}-}\right.\\
\notag
\left.{
-\frac{\mu^{j3}_{m'} \mu^{j4}_{m}}{\omega_3 - E_m - E_{m'} + E_{m'}}-
\frac{\mu^{j3}_{m} \mu^{j4}_{m'}}{\omega_3 - E_m - E_{m'} + E_{m}}
}\right]
\end{gather}
This signal vanishes identically despite the Pauli induced scattering making it a measure of the screened Coulomb interaction.

\end{appendix}

\end{document}